\documentclass[final,5p,times, twocolumn]{elsarticle}

\usepackage{amssymb}

\usepackage{lineno}
\usepackage{setspace}

\usepackage[colorinlistoftodos]{todonotes}

\journal{Colloids and Surface A}

\begin{document}

\begin{frontmatter}



\title{3D simulations of wet foam coarsening evidence a self similar growth regime}


\author[IFUFRGS]{Gilberto L. Thomas}

\author[Indiana]{Julio M. Belmonte\fnref{EMBL}}

\author[Diderot]{Fran\c{c}ois Graner}

\author[Indiana]{James A. Glazier}

\author[IFUFRGS,Indiana,INCT]{Rita M.C. de Almeida\corref{cor1}}

\address[IFUFRGS]{Instituto de F\'\i sica, Universidade Federal do Rio Grande do Sul\\ Av. Bento Gon\c{c}alves 9500, C.P. 15051 - 91501-970 Porto Alegre, RS, Brazil }
\fntext[EMBL]{Present address: European Molecular Biology Laboratory Heidelberg, Meyerhofstr. 1, 69117 Heidelberg, Germany}
\cortext[cor1]{Email: rita@if.ufrgs.br, Tel: +555133086521, Fax: +5533087286}
\address[Indiana]{Biocomplexity Institute and Department of Physics, Indiana University Bloomington, Bloomington, Indiana, 47405, United States of America}
\address[Diderot]{Mati\`ere et Syst\`emes Complexes, Universit\'e Paris Diderot, CNRS UMR 7057, 10 rue Alice Domon et L\'eonie Duquet, F-75205 Paris Cedex 13, France}
\address[INCT]{Instituto Nacional de Ci\^encia e Tecnologia - Sistemas Complexos\\ Av. Bento Gon\c{c}alves 9500, C.P. 15051 - 91501-970 Porto Alegre, RS, Brazil}

\begin{abstract}

In wet liquid foams, slow diffusion of gas through bubble walls changes bubble pressure, volume and wall curvature. Large bubbles grow at the expenses of smaller ones. The smaller the bubble, the faster it shrinks. As the number of bubbles in a given volume decreases in time, the average bubble size increases: \textit{i.e.} the foam coarsens. During coarsening, bubbles also move relative to each other, changing bubble topology and shape, while liquid moves within the regions separating the bubbles. Analyzing the combined effects of these mechanisms requires examining a volume with enough bubbles to provide appropriate statistics throughout coarsening. 
Using a Cellular Potts model, we simulate these mechanisms during the evolution of three-dimensional foams with wetnesses of $\phi=0.00$, $0.05$ and $ 0.20$. 
We represent the liquid phase as an ensemble of many small fluid particles, which allows us to monitor liquid flow in the region between bubbles. The simulations begin with $2 \times 10^5$ bubbles for $\phi = 0.00$ and $1.25 \times 10^5$ bubbles for $\phi = 0.05$ and $0.20$, allowing us to track the distribution functions for bubble size, topology and growth rate over two and a half decades of volume change. All simulations eventually reach a \textit{self-similar} growth regime, with the distribution functions time independent and the number of bubbles decreasing with time as a power law whose exponent depends on the wetness.
\end{abstract}

\end{frontmatter}
\doublespacing



\section{Introduction}

Liquid foams include soap froths and food and industrial foams, as well as many liquid-phase pollutants. Foams consist of gas bubbles surrounded by a continuous liquid phase. Bubble walls are very thin; their thickness strongly depends on the intensity of mutual repulsion between surfactant-loaded gas-liquid interfaces and only weakly on foam \textit{wetness} ($\phi$={liquid volume}/{total volume}) below a critical wetness at which the foam transforms into a \textit{bubbly liquid}. When three bubble walls meet, they form edges, known as \textit{Plateau borders}, which are tube-like structures with a cross section which is roughly a triangle with its straight sides replaced by concave circular arcs. Wet foams and dry foams both have bubbles separated by thin walls, the difference being that wetter foams have larger Plateau borders \cite{Weaire2001,Cantat2010}.  

\begin{figure}
\centering
\includegraphics[width=0.45\textwidth]{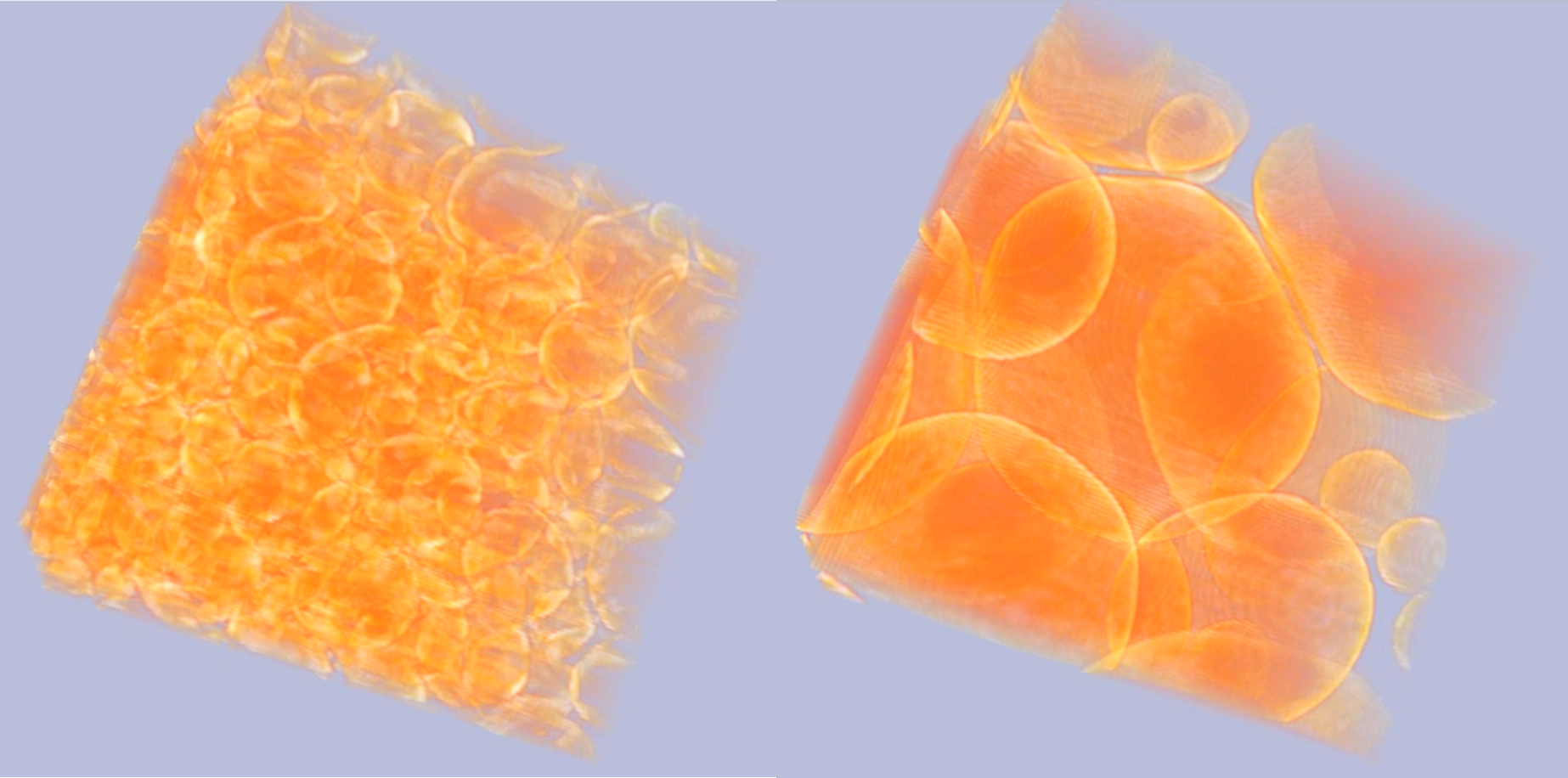}
\caption{Snapshots of a 3D foam-coarsening experiment. First (left) and last (right) images of a 3D movie obtained by \cite{Lambert2007}, in a foam with $\phi=0.17$.}
\label{experiment}
\end{figure}

Slow diffusion of gas through bubble walls and Plateau borders, due to the Laplace pressure difference, changes bubble pressure, volume and wall curvature. Bubbles, Plateau borders and walls move and relax towards mechanical equilibrium, while liquid flows inside the Plateau borders and walls as well. Bubbles larger than the mean bubble size (or bubbles with more than the mean number of faces for dry foams) grow at the expenses of smaller (fewer-faced) ones. Smaller (fewer-faced) bubbles shrink more and more quickly, and eventually disappear. The number of bubbles decreases in time, and the average bubble size increases: the foam \textit{coarsens}~\cite{Weaire2001,Cantat2010,Saint-Jalmes2006} (figure \ref{experiment}).  

X-ray tomography experiments \cite{Lambert2010} and numerical simulations \cite{Thomas2006,Streitenberger} have found that dry foams reach a self-similar growth regime where the topology and relative bubble size distributions are steady while the mean bubble radius increases as $t^{1/2}$, as expected from dimensional analysis~\cite{mullins86}. Hence, the mean bubble volume increases as $t^{3/2}$ and the number of bubbles decreases as $t^{-3/2}$.  
In the opposite limit, in a bubbly liquid, gas flows from small bubbles to the liquid phase, where it diffuses, to eventually flow back into larger bubbles due to Laplace pressure difference, now between the bubbles' internal pressures and the gas saturation pressure in the liquid. This \textit{Ostwald ripening} dynamics also generates a self-similar growth regime, where the relative bubble size distribution is constant, and the mean bubble radius grows as $t^{1/3}$ \cite{Baldan2002}. Hence, the mean bubble volume grows as $t$ and the number of bubbles decreases as $t^{-1}$. 

To investigate coarsening for foams of intermediate wetnesses, Isert, Maret and Aegerter \cite{Isert2013} levitated an aqueous foam in a magnetic field: the water's diamagnetism counterbalanced gravity, so the wetness remained uniform throughout the foam. At long times, for foams with wetness less than $\phi=0.25$ the mean bubble radius grew as $t^{1/2}$ as in a dry foam, for foams with wetness greater than $\phi=0.35$ the mean bubble radius grew as $t^{1/3}$ as in a bubbly liquid, and the mean bubble radius growth exponent decreased from $1/2$ to $1/3$ between $\phi=0.25$ and $\phi=0.35$. Determining whether power law growth in bubble size at these wetnesses reflected self-similar growth with constant relative bubble size distributions would require determining individual bubble properties within the levitated foam. Such a measurement would be quite difficult, though it might be possible using \textit{e.g.} X-ray tomography ~\cite{Lambert2007}.

Numerical simulations provide access to properties of individual bubbles (volume, face-number and edge-number distributions, and their spatial correlations) with arbitrary spatial and temporal resolution. Complicating such simulations are the multiple mechanisms, length and time scales affecting coarsening. Slow volume changes in bubbles cause the gradual increase of Plateau border cross section. Both bubble and Plateau-border growth can induce topological transformations in bubble and Plateau-border adjacency, such as neighbors swapping (\textit{T1}), bubble disappearance (\textit{T2}), or Plateau-border fusion, which can cause further rapid cascades of rearrangement. These rearrangements, in turn, change the slower dynamics of gas exchange (between neighboring bubbles, and between bubbles and the liquid phase) and flow in Plateau borders. Length scales include the thickness of the thin walls (which can range from hundreds of nanometers to many microns), the size of the Plateau borders (which can range from hundreds of microns to millimeters), the lengths of Plateau borders and the radii of bubbles (which can range from sub-millimeter to centimeters) depending on the foam, wetness and  degree of coarsening. The close feedback among all these mechanisms and scales makes simulation of wet foam dynamics challenging~\cite{Saye2013}. 

\begin{figure}
\centering
\includegraphics[width=0.25 \textwidth]{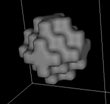}
\caption{Representation of the grid neighborhood for Potts energy calculations, up to the 5$^{th}$ layer, summing over 56 neighbors.}
\label{julio}
\end{figure}

The Cellular Potts model (\textit{CPM}) can simulate the structure of foams in mechanical equilibrium, with a resolution (number of pixels per bubble)  which matches or exceeds that of experimental images in two or three dimensions.
The CPM can also simulate the quasi-static dynamics of foam coarsening, as slow gas diffusion leads to a succession of mechanically equilibrated foam structures. The CPM correctly reproduces bubble rearrangements and disappearance. 

The CPM developed from Potts model simulations of coarsening in metallic polycrystals \cite{Anderson1989}, because the mass flux between bubbles and crystals both obey Fick's law (a local dynamics which produces the same growth dynamics as the Young-Laplace law for bubbles). Such CPM simulations agree in almost all respects with experimental observations of dry liquid foams \cite{Glazier1993}. However, in the CPM, mass flow between bubbles and surface-tension-driven relaxation of wall shape occur over similar time scales, while the former is much slower than the latter in almost all liquid foams. Extending the CPM method by interleaving steps in which mass is free to diffuse between bubbles with steps in which we constrain the bubble volumes to their current values but allow the bubble walls to relax allows us to reproduce the experimental hierarchy of time scales. The more relaxation steps per diffusion step, the faster the  relaxation rate relative to the diffusion rate \cite{Thomas2006}.

CPM simulations of 3D dry foam coarsening \cite{Thomas2006,Streitenberger} and 2D wet foam coarsening \cite{Fortuna2012} both reach self-similar growth regimes. Here we compare 3D CPM simulations (with additional shape relaxation steps) of foams with wetnesses of $\phi=0.00$, $0.05$ and $0.20$. All three reach self-similar growth regimes with enough bubbles to yield statistically significant distribution functions.

\section{Methods}
\subsection{Foam structure}

Like experimental images, CPM 3D foam simulations represent space as a cubic grid of voxels, and each bubble as a connected set of voxels with particular properties. We employ a space of $V_T= 200^3$ voxels with periodic boundary conditions, initially filled with $2 \times 10^5$ bubbles for $\phi = 0.00$ and $1.25 \times 10^5$ bubbles for $\phi = 0.05$ and $0.20$. Each voxel, located at a position $\vec{r}$, has a label $S$, corresponding to a bubble or portion of liquid. Defining the liquid phase as a single domain would allow liquid to move instantaneously within the foam, effectively giving it zero inertia and viscosity. To ensure local conservation of liquid volume and to keep the liquid velocity finite, we follow Fortuna {\em et al.} \cite{Fortuna2012}, and subdivide the liquid phase into numerous small fluid particle subdomains (see figure S1 in the supplementary materials).
  
The CPM energy function includes a surface energy and a volume constraint:   
\begin{equation}
\label{energy}   
E = \sum_{\vec{r}} \sum_{\vec{v}(\vec{r})} J(S_{\vec{r}}, S_{\vec{v}}) + \sum_{S} \lambda(S)\;\left(V_{S} - V^{target}_{S}\right)^2
\end{equation}
where the sum over $\vec{v}(\vec{r})$ extends through the $5^{th}$ neighbors around the voxel at $\vec{r}$ (56 neighbors), to reduce lattice-anisotropy effects \cite{Holm1991}, and is only evaluated for neighboring voxels with distinct labels $S$. Figure \ref{julio} illustrates the grid neighborhood in the energy calculation. The second term in the r.h.s. of Eq.~(\ref{energy}) constrains the volume of the labeled region $S$, $V_{S}$, around its target volume, $V^{target}_{S}$, when $\lambda(S)$, which is the compression modulus (inverse compressibility) of the material, is $ >0$. 
 
\begin{figure}
\centering
\includegraphics[width=0.43 \textwidth]{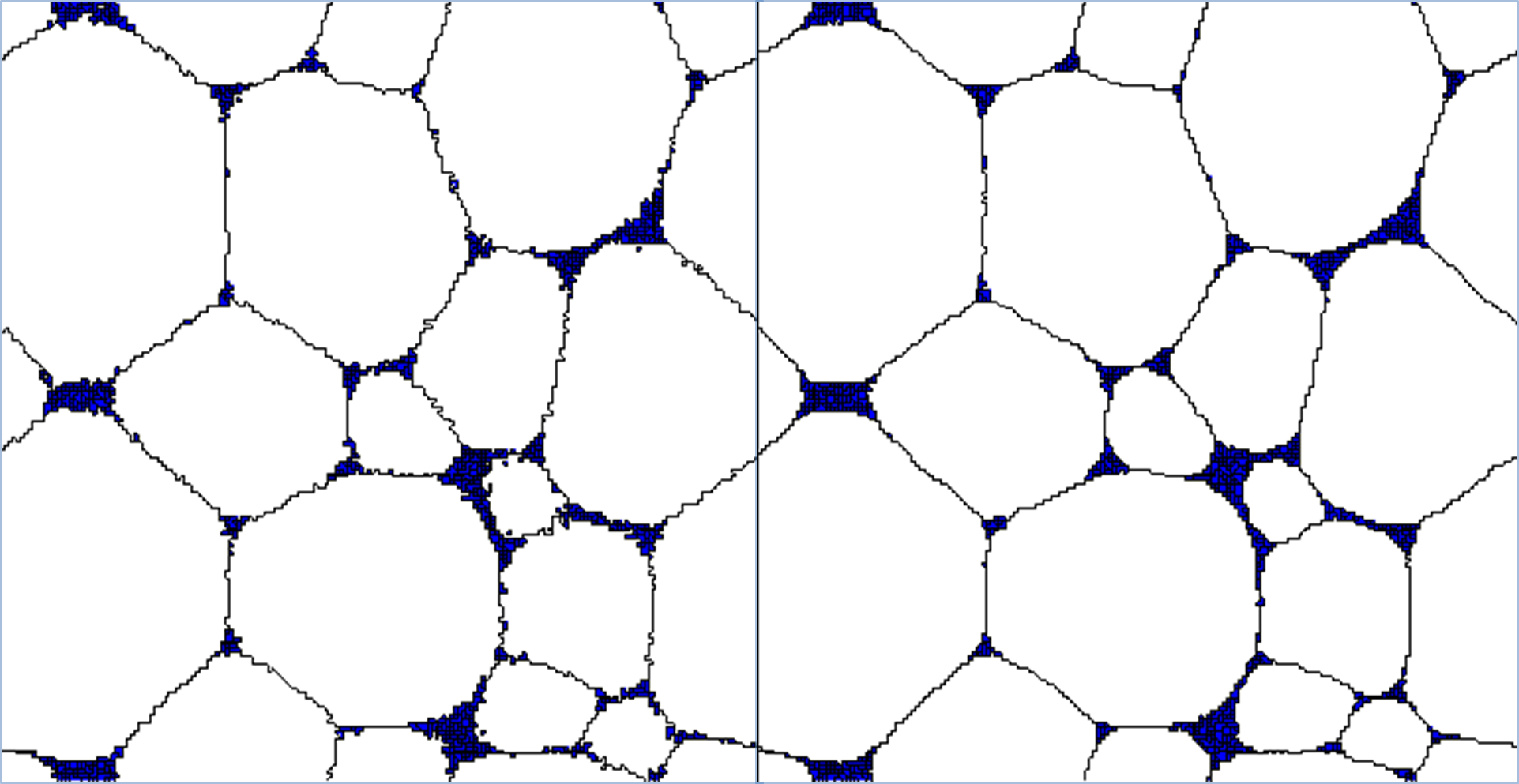}
\caption{Cross-sections through a $200^3$ voxels 3D CPM simulation of a liquid foam with periodic boundary conditions, for $\phi \simeq 0.05$ after $5600$ MCS. Left: without interface relaxation. Right: with interface relaxation.}
\label{anneal}
\end{figure}

\begin{figure}
\centering
\includegraphics[width=0.47 \textwidth]{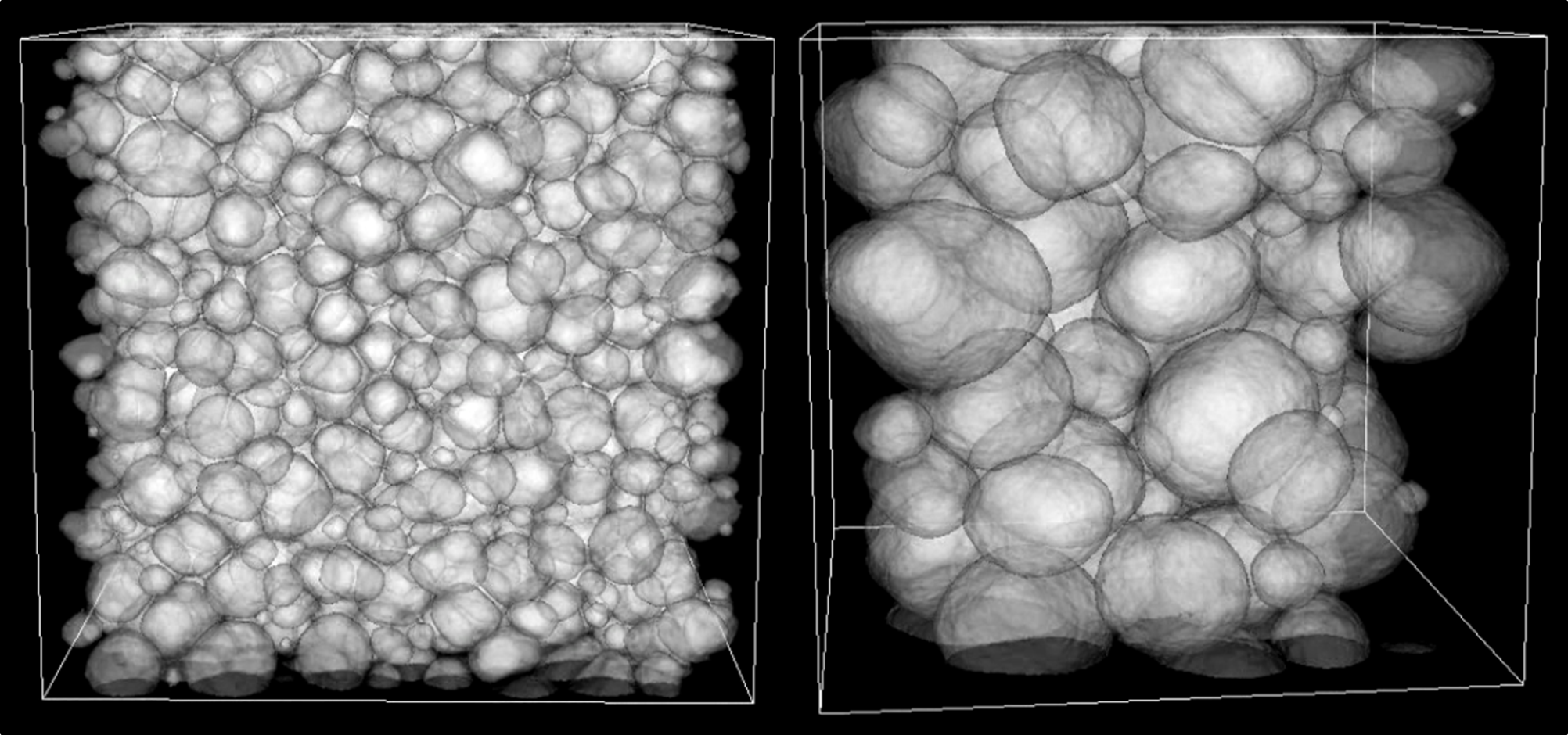}
\caption{Images of a CPM simulation of a liquid foam with $\phi = 0.20$, run in a grid with $300^3$ voxels. Initial (left) and final (right) states.}
\label{threedimage}
\end{figure}

While an ``ideal" fluid particle would be negligibly small and perfectly incompressible, we assign our fluid particles a $V^{target}=7$ voxels so they are much smaller than typical bubbles, but large enough to move freely (see figure S2 in the supplementary materials for the effect of varying the target volume of the fluid particles, which affects coarsening dynamics). The compression modulus $\lambda_{liq}=10$ is not critical, since coarsening dynamics are robust under moderate variation of this parameter (see figure S3 in the supplementary materials). However, making $\lambda_{liq}$ too large (\textit{i.e}. the compressibility too small) leads to non-physical ``freezing" of the liquid in the Plateau borders.

Simulations begin with a number of fluid particles randomly distributed over the grid. The foam wetness $\phi$ is simply the number of voxels in fluid particles (controlled by their number and target volume) divided by the lattice size. We fill the rest of the grid with bubbles with initial target volumes randomly drawn from a log-normal, normal or bidisperse probability distribution.

The first term of the r.h.s. of Eq.~(\ref{energy}) defines the interfacial energy between the different types of regions:
\begin{eqnarray}
\label{jotas}
J(liq,liq) & = & 0.10, \nonumber \\
J(liq,gas)=J(gas,liq) &=&  1.00, \nonumber \\
J(gas,gas) & =& 1.99.   
\end{eqnarray}
$J(liq,liq)$ would be zero in a real liquid, but a small positive value helps to maintain the fluid particles as connected domains without significantly affecting their dynamics. $J(liq,gas)$ is the energy per unit area of the gas-liquid interface and $J(gas,gas)$ is of the order of $ 2J(liq,gas)$ because a wall separating two bubbles consists of two gas-liquid interfaces. If $J(gas,gas)$ were strictly equal to $2J(liq,gas)$, the fluid particles would spread in the walls as in a bubbly liquid. We set $J(gas,gas)$ slightly lower than $2J(liq,gas)$ to account for the positive disjoining pressure ($2J(liq,gas)-J(gas,gas)$) between the soap films in the walls, which serves to keep the walls thin, as observed in experiment. The greater the disjoining pressure, the greater the fraction of liquid that localizes in the Plateau borders and the less that penetrates into the bubbles walls. As a result, we expect coarsening to be faster for higher disjoining pressures. For the range $2J(liq,gas)-J(gas,gas)=0.1$, $0.01$ and $0.001$, the simulation results are robust and the liquid's spatial distribution is essentially the same (see figure S4 in the supplementary materials). 

\begin{figure}
\centering
\includegraphics[width=0.43\textwidth]{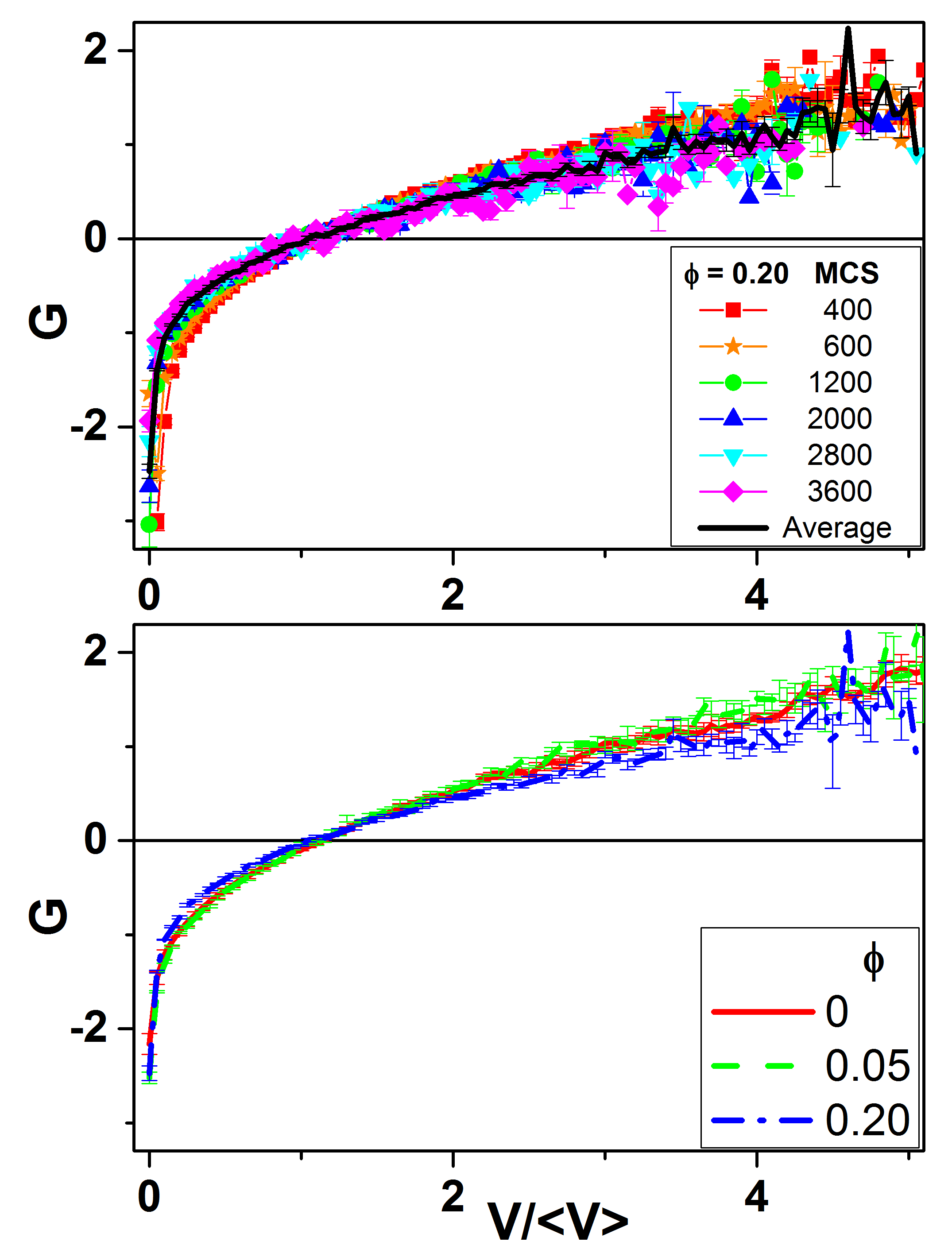}
\caption{Growth rate. Upper panel: Growth rate as function of the normalized bubble volume $V/\left\langle V \right\rangle$  for simulations starting from a log-normal bubble target-volume distribution with $\phi = 0.20$ (top). The black line shows the average over 10  simulation runs for 31 time points after the onset of self-similar growth. Lower panel: Self-similar growth rate as a function of the normalized bubble volume $V/\left\langle V \right\rangle$ for different wetnesses. While similar, these growth rates are not identical, showing a significant dependence on $\phi$.}
\label{growthV}
\end{figure}

The wall thickness and Plateau-border shape depend on the interaction range ($\vec{v}(\vec{r})$ in Eq.~(1)) and the size of the fluid particles.

\subsection{Foam dynamics}

Configurations change stochastically in the CPM. We randomly pick a pair of first neighbor voxels with different labels and calculate the current energy, $E_i$, and the energy after copying the label $S$ from the first to the second voxel, $E_f$. If $\Delta E=E_f-E_i \le 0$ we always make the copy. If $ \Delta E>0$ we make the copy with probability $ \exp{(-\Delta E / T)}$, where $T$ is a Boltzmann-like fluctuation amplitude which we choose just large enough to overcome lattice pinning effects (the macroscopic fluctuation amplitude in a real foam would be close to $0$, but real foams do not experience lattice pinning). The simulation unit which corresponds to time is the \textit{Monte Carlo Step} (\textit{MCS}): $1$ MCS consists of $N$ voxel-copy attempts, where $N$ is the total number of voxels in the grid.

Like all models, the CPM's simplification of reality may introduce artifacts. The CPM models mass transfer as voxel transfer. The transfer of mass between two bubbles in contact across a thin wall obeys Fick's law in both CPM and experiment, so the key question is the rate of transfer as a function of parameters. In both experiment and simulation, we expect the transfer of mass across the Plateau borders to be slower than across thin walls. However, the dependence of the rate of transfer of bubble mass between bubbles via the Plateau borders depends less obviously both on model parameters and geometry and is not simple to separate from transfer across thin walls, so both the rate and functional form of the transfer might differ from experiment. Prior work has shown that 2D CPM simulations of wet foams using the current method reproduce the bulk growth behaviors and distribution functions seen in experiments, including their dependence on foam wetness \cite{Fortuna2012}. While such agreement shows that the functional form of the mass transfer via Plateau borders must be similar in CPM and wet foams, the agreement does not demonstrate that the functional forms are identical, nor how the rate depends on model parameters. In particular, for foams composed of a gas with fairly low solubility in the liquid phase, we expect gas transport via walls to be much faster than via Plateau borders. In our current CPM simulations, the interaction range is a significant fraction of the typical Plateau-border size, thus we expect the transport rates via walls and Plateau borders to be more similar and the difference in growth rate between dry and wet foams to be less pronounced in the simulations than in an experiment with a foam with a low solubility gas phase. In the near future, we plan to examine the explicit relationship between Plateau border shape, size and model parameters, which will allow us to control the relative transport rates explicitly.

We first relax the starting configuration for $200$ MCS, to allow the bubble and liquid volumes and geometries to adjust to generate a foam with the initially specified area distribution, then begin the simulation proper. To ensure that the equilibration of wall and Plateau-border shapes is much faster than the rate of mass diffusion, after each 20 MCS of coarsening steps we run at least 10 MCS of relaxation steps, continuing until the number of accepted voxel copies during a relaxation MCS is zero or less than $1 \%$ of the number of accepted voxel copies during the previous relaxation MCS. Because mass transfer controls the time-scale of coarsening in experiment, with relaxation occuring simultaneously, the simulation time $t$ in our graphs reflects the number of coarsening steps only. During coarsening steps we set $\lambda_{gas}=0$ so bubble volumes are free to vary and $T=8$ for $\phi=0$, and $T=25$ for $\phi>0$ since the volume constraint on the fluid particles increases their pinning to the grid. During relaxation steps, we set $T=0$, $\lambda_{gas}=7$, and $V^{target}_S$ for each bubble equal to the bubble's volume at the beginning of the step. The volume constraint prevents mass transfer between bubbles, while allowing boundaries to move to minimize their interface energy.

\begin{figure}
\centering
\includegraphics[width=0.43\textwidth]{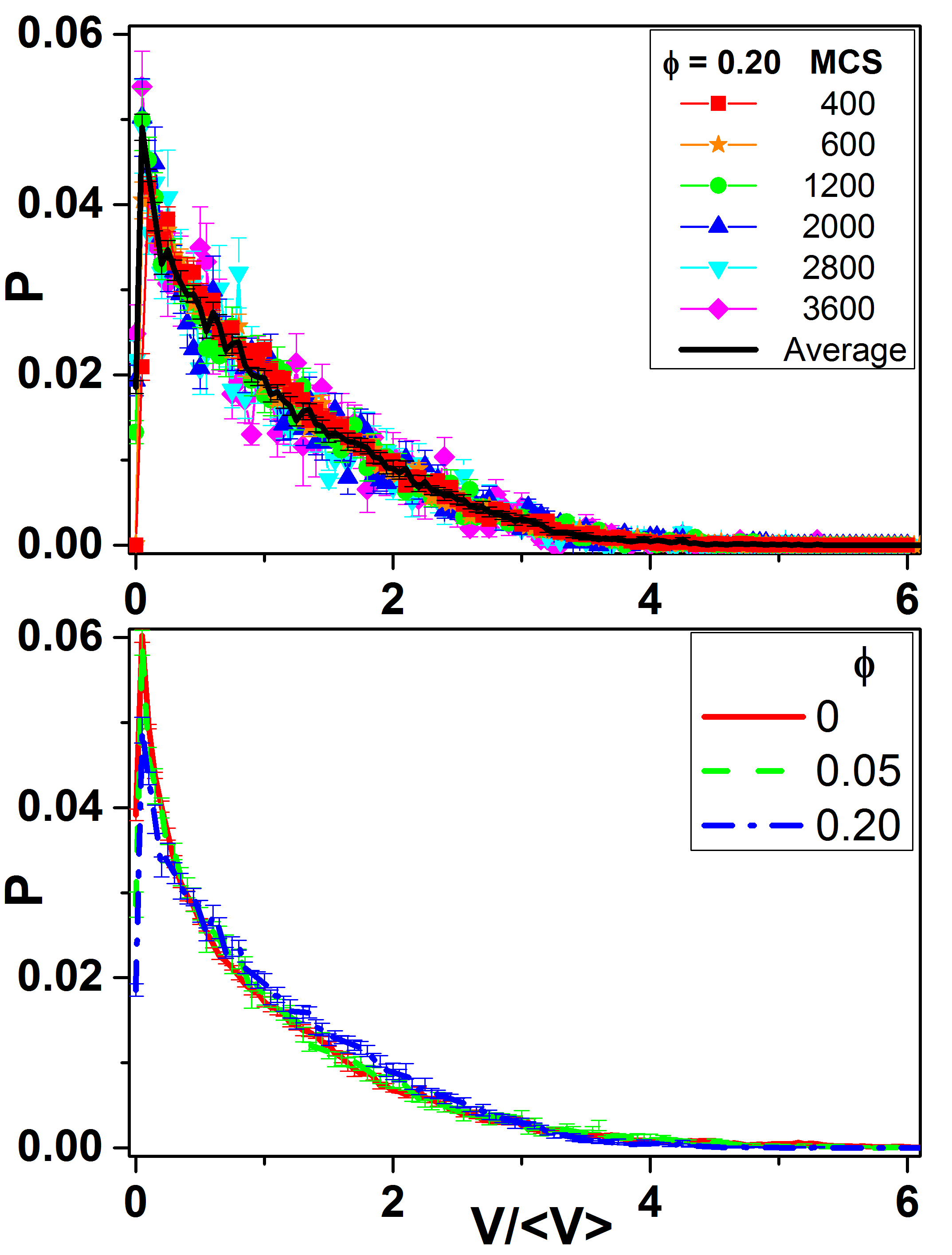}
\caption{Bubble size distribution. Upper panel: Volume distributions as function of $V/\left\langle V \right\rangle$ for simulations starting from a log-normal bubble volume distribution and $\phi = 0.20$ (top). The black line shows the average over 10 simulation runs for 31 time points after the onset of self-similar growth. Lower panel: Self-similar volume distributions for different wetnesses. As for the growth rates, the distributions are similar but statistically distinct. }
\label{PV}
\end{figure}

The simulations use the CompuCell3D environment \cite{Swat2012}, publicly available at http://www.compucell3d.org/. We ran the simulations on an Intel Core I7- 3700K processor. We ran 10 replicas for each set of simulation parameters, starting from a log-normal volume distribution for each set of parameters, with $\phi=0.00$, $0.05$, and $0.20$. Each simulation replica took from 2 to 7 days to run, depending on the number of simulated objects (bubbles plus fluid particles), which varies with wetness. Because CompuCell3D limits the total number of objects in a simulation, we chose the size of the grid to respect that limit.

\section{Results}

Figure \ref{anneal} shows 2D sections of a simulated 3D foam before (left) and after (right) a set of relaxation steps. Relaxation smoothes the boundaries while preserving bubble volumes. Figure \ref{threedimage} shows 3D images of a larger simulation for $\phi=0.20$, run on a $300^3$ grid: boundaries are smooth and the foam texture closely resembles that in the experiments in Fig.\ref{experiment}.

\begin{figure}
\centering
\includegraphics[width=0.43\textwidth]{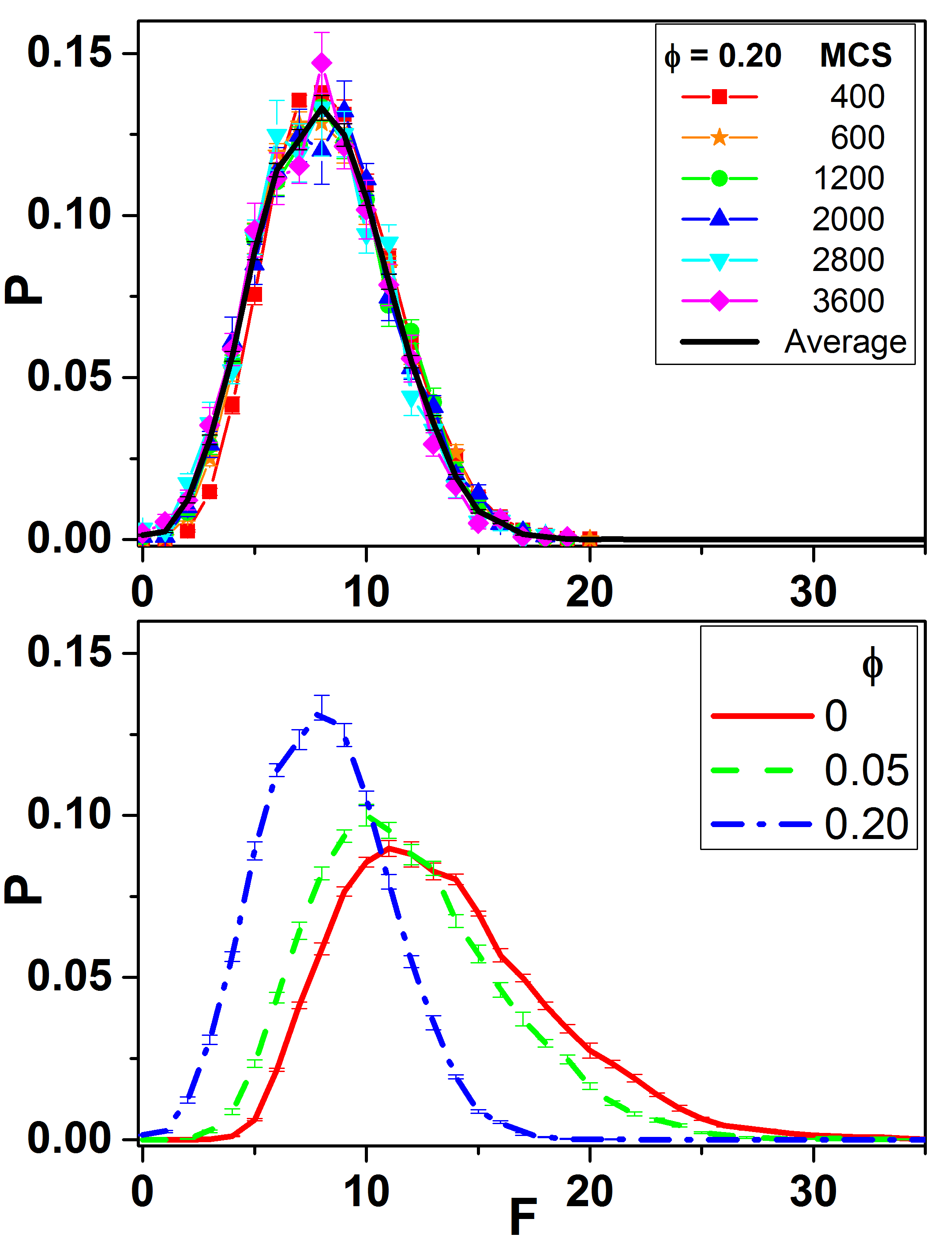}
\caption{Bubble face-number distribution. Upper panel: Face-number distributions  for simulations starting from a log-normal bubble volume distribution for $\phi = 0.20$ (top). The black line shows the average over 10 simulation runs for 31 time points after the onset of self-similar growth. Lower panel: Self-similar bubble face-number distributions for different wetnesses. }
\label{PF}
\end{figure}

To check whether the foam reached a self-similar growth regime  we compared the bubble growth rate, $G$, as a function of the normalized bubble volume $V/\left\langle V \right\rangle$ (figure \ref{growthV}) at different times. The growth rates are identical within error for $t>600$ MCS for every wetness value (upper panel for $\phi = 0.20$, data not shown for $\phi =0.00, 0.05$). The lower panel shows the self-similar growth rates (averaged over 10 runs and 31 times) for $\phi=0.00$, $0.05$, and $0.20$. Figure S5 in the supplementary materials shows a detail of this figure: the growth rates for $\phi=0.20$ significantly differ from those in the drier foam simulations.

\begin{figure}
\centering
\includegraphics[width=0.43\textwidth]{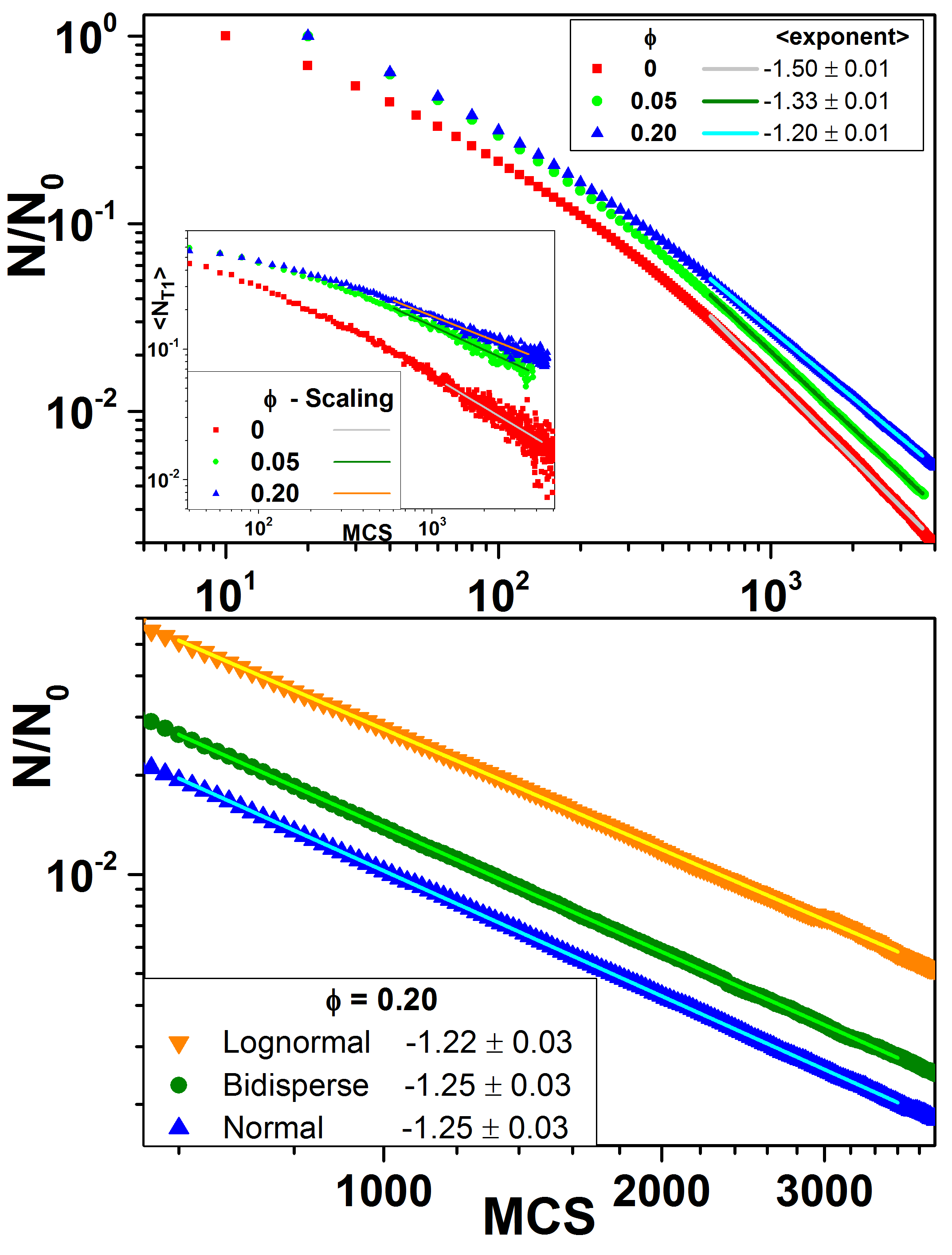}
\caption{Bubble number vs time. Upper panel: Change in the number of bubbles for wetnesses $\phi = 0.00$, $0.05$ and $0.20$, starting from a log-normal distribution of bubble volumes. After a transient period of roughly $600$ MCS, the number of bubbles and the rate of T1s (inset) both decrease as a power of $t$. Lower panel: Dependence of the fractional decrease in the number of bubbles for different initial volume distributions with $\phi=0.20$. Averages and standard deviations calculated from 10 simulation replicas per data point.}
\label{numBubbles}
\end{figure}

Figures \ref{PV} and \ref{PF} show the distributions of normalized bubble volumes and face number. Both distributions are time invariant for $t > 600$ MCS for all wetnesses. The lower panels compare the average distributions for simulations with $\phi=0.00$, $0.05$, and $0.20$. The normalized volume distributions are similar, though statistically distinct (see figure S6 in the supplementary materials). The face-number distributions differ dramatically for foams with different wetnesses.

The upper panel in figure \ref{numBubbles} shows log-log plots of the decrease in number of bubbles starting from initial configurations with a log-normal distribution of bubble volumes. The straight line behavior observed  for all wetnesses for $t>600 $ MCS over roughly one decade is compatible with a power law, which is a requirement for a self-similar growth regime \cite{mullins86}. Fits for $600$ MCS $\le t \le 3500$ MCS yield volume scaling exponents of $1.50$, $1.33$, and $1.20 \pm 0.01$, for $\phi = 0.00, 0.05$ and $0.20$ respectively (corresponding to radius scaling exponents of $0.5$, $0.44$ and $0.4$). For $t > 600$ MCS, the number of bubble rearrangements $N_{T1}$ per MCS also decreases as a power of time (figure \ref{numBubbles}, inset in the upper panel). The lower panel in figure \ref{numBubbles} shows log-log plots of the decrease of the number of bubbles for $\phi=0.20$, starting from initial configurations with normal, log-normal and bidisperse bubble volume distributions. The growth exponent is independent of the initial volume distribution. Figure S7 in the supplementary materials shows the self-similar distributions for number of faces for $\phi=0.20$ and different initial distributions: the overlapping of these functions is a further evidence for the scaling regime.

\section{Discussion and conclusions}

So far, experiments and simulations have investigated self-similar growth primarily in dry foams \cite{Lambert2010,Thomas2006,Streitenberger}. Together, the growth rates in figure \ref{growthV} and distribution functions in figures \ref{PV} and \ref{PF}, and the power laws seen in figure \ref{numBubbles}, show that the foams reach a self-similar growth regime after a transient of $\approx 600$ MCS for all wetnesses and all three initial volume distributions.

Radius scaling exponents in experimental foams depend on the wetness, dropping from 1/2 to 1/3 as $\phi$ goes from zero to one. In our simulations, the radius scaling exponent for $\phi=0.20$ is significantly smaller than in the experiments of Isert \cite{Isert2013} but similar to that in Fortuna \textit{et al.} \cite{Fortuna2012}. This difference between experiments, as well as between experiment and simulation, may depend on the relative transport rates in thin walls and Plateau borders, which we will investigate in future work.

\section{Acknowledgements}

RMCdA acknowledges the kind hospitality at the Biocomplexity Institute, Indiana University, Bloomington. This Work was partially supported by Brazilian agencies CAPES and CNPq-FAPERGS (PRONEX 10/0008-0), and the USA National Institute of Health grants UO1 GM111243, UO1 GM076692, and RO1 GM077138.

\end{document}